\documentclass[12pt]{iopart}

\usepackage{epsfig,iopams}
\usepackage{graphicx}  

\let\la=\lesssim     
\let\ga=\gtrsim

\newcommand{\arcsec}{\mbox{\ensuremath{^{\prime\prime}}}}
\newcommand{\sun}{\ensuremath{\odot}}
\newcommand{\aap}{{\it Astron.~Astrophys.}}

\newcommand{\apj}{{\it Astrophys.~J.}}
\newcommand{\apjs}{{\it Astrophys.~J.~Supp.}}
\newcommand{\aj}{{\it Astron.~J.}}
\newcommand{\mnras}{{\it Mon.~Not.~R.~Astron.~Soc.}}
\newcommand{\prl}{{\it Phys.~Rev.~Lett.}}

\begin{document}

\title[A comparison of the strong lensing properties of the S\'ersic and the NFW profiles]{A comparison of the strong lensing properties of the S\'ersic and the NFW profiles}


\author{\'A. El\'iasd\'ottir$^{1}$ and O. M\"oller$^{2}$}
\address{$^{1}$Dark Cosmology Centre, Niels Bohr Institute, Univ. of Copenhagen, Juliane Maries Vej 30, Copenhagen 2100, Denmark (ardis@dark-cosmology.dk)}
\address{$^{2}$Max-Planck-Institut f\"ur Astrophysik, Karl-Schwarzschild-Strasse 1, D-85741 Garching, Germany (ole@mpa-garching.mpg.de)}


\label{firstpage}

\begin{abstract}
We investigate the strong lensing properties of the S\'ersic profile as
an alternative to the NFW profile, focusing on applications to lens modelling of clusters.   Given an underlying S\'ersic dark matter profile, we study whether an NFW profile can provide an acceptable fit to strong lensing constraints in the form of single or multiple measured Einstein radii.  We conclude that although an NFW profile that fits the lensing constraints can be found in many cases, the derived parameters may be biased. In particular, we find that for $n\sim2$, which corresponds to massive clusters, the mass at $r_{200}$ of the best fit NFW is overestimated (by a factor of $\sim 2$) and the concentration is very low ($c\sim2$).   The differences are important enough to warrant the inclusion of  S\'ersic profile for future analysis of strong lensing clusters.
\end{abstract}


\section{Introduction}
According to $\Lambda$CDM simulations, dark matter is expected to
dominate baryonic matter on both galactic and cluster scales.
Observations, including X-ray studies, gravitational lensing and
rotational velocity curve measurements, also suggest that the luminous matter is
only a fraction of the total matter in the Universe, although it dominates the dark matter in the innermost regions of galaxies and clusters.  Understanding
the nature of dark matter is one of the key challenges in
modern astrophysics.  In particular, a knowledge of the spatial
distribution of the dark matter is crucial for understanding its
interplay with the baryonic matter and gaining insight into its
nature.

The dark matter distributions for halos from $\Lambda$CDM simulations
are usually described by the Navarro, Frenk, \& White (NFW) profile
\cite{nfw1}, but recently \cite{navarro2004} found a better fit using a S\'ersic profile for the 3D density distribution.  That work was expanded in \cite{merritt} which found that the S\'ersic distribution also provides a good fit to the 2D distribution of dark matter halos and found the 
deprojected S\'ersic profile to give the best fit to the 3D distribution.  It is intriguing that the S\'ersic law, which is often fit to
the 2D luminosity profiles of elliptical galaxies \cite{sersic, ciotti1991, caon,graham2003}, should also describe the surface
density profiles of dark matter halos from simulations.  In fact, work
by e.g. \cite{hayashi2004} suggests that the S\'ersic profile
provides a better fit in particular to the inner regions of dark
matter halos, where the DM density profile affects the kinematics of
the central galaxies in these halos more strongly. As
lensing can probe the surface density profiles of galaxies (see e.g., \cite{rusin,ferreras}) and clusters (see e.g., \cite{kneib,broadhurst}), especially in their central regions, it
is of interest to compare the lensing properties of the S\'ersic and
the NFW profiles; does the use of an NFW profile for lensing mass
reconstructions introduce a significant bias given an actual S\'ersic
lens profile, or are the two essentially indistinguishable for lensing
mass reconstructions?

The question of the slope for the inner profile in lens systems like
galaxies and clusters has been addressed by several authors
before \cite{oguri2004, sand2004}. Others have
predicted and/or attempted to reconstruct the slope of lensing systems
from the observed positions and magnifications of multiply-imaged
sources \cite{broadhurst,koopmans2006}. In this work, we take a very theoretically motivated approach:
given 
either an NFW or S\'ersic profile for dark mass
distributions, how would their lensing properties differ? We thereby
neglect the fact that in actual lens systems, the lensing mass also
includes a baryonic component that is parameterised in a different
way. Even if the stellar mass component of early-type galaxies can
also be described by a S\'ersic profile, the scale length and central
densities are very different than those found in simulations for the 
S\'ersic profiles for the dark matter component. The work presented here
therefore only describes the {\it difference} in lensing properties
between NFW and S\'ersic profiles if the total mass distribution can be
described by a single S\'ersic profile. For cluster systems, the dark matter
component is a more significant contributor to the lensing properties
and the baryonic component plays a smaller role than for galaxy lens
systems. In this sense the results in our paper are more readily
compared with current lensing constraints on the dark matter profiles
of clusters. The analytical lensing relations derived in Section 2
are, however applicable to all lens systems that can be described by
a S\'ersic profile, irrespective of the scale and whether the matter
is baryonic or dark.

Throughout this paper we assume standard $\Lambda$CDM cosmology with $\Omega_\Lambda=0.70$, $\Omega_m=0.30$ and $H_0 = 70$ km s$^{-1}$ Mpc$^{-1}$ ($h=0.7$).  For numerical lensing calculations we place the lens at a redshift of $z=0.3$ (a realistic redshift for both galaxy and cluster lenses) and the source at $z=10$ (chosen to approximate a source at $z=\infty$, although we note that for any $z\ga1$ the results are similar).  For the given cosmology, $1\arcsec$ corresponds to $4.45$~kpc at $z=0.3$.  The paper is organised as follows: We start by introducing the NFW and the S\'ersic profiles in the remainder of Section 1.  In Section 2 we derive and discuss the magnification properties of the S\'ersic profile.  In Section 3 we study and discuss the differences in the lensing properties of a S\'ersic surface matter density and a corresponding best fitting NFW.  We study whether applying an NFW fit to an underlying S\'ersic profile leads to a bias in the mass and concentration parameter and discuss the implications our results could have on modelling of observed lenses.  We summarise our conclusions in Section~\ref{sec:summary}.

\subsection{The NFW profile}
The NFW profile \cite{nfw1,nfw2} has been extensively used to fit $\Lambda$CDM halos from simulations and in 3D is given by:
\begin{equation}
\label{eq:ro_nfw}
\rho_{\mathrm{nfw}}=\frac{\delta_c \rho_c}{(r/r_s)(1+r/r_s)^2}
\end{equation}
where $r_s=r_{200}/c_{200}$ is the (3D) scale radius, $c_{200}$ is a dimensionless number called the concentration parameter, $\rho_c$ is the critical (3D) density at the redshift of the halo, $r_{200}$ is the radius inside which the mass density of the halo equals $200 \rho_c$ and $\delta_c$ is the characteristic over-density for the halo given by
\begin{equation}
\delta_c=\frac{200}{3}\frac{c_{200}^3}{\ln(1+c_{200})-c_{200}/(1+c_{200})}.
\end{equation}

The 2D projection of the NFW profile is frequently used in the modelling of gravitational lenses \cite{golse,williams, broadhurst,wayth}.  It is found by integrating equation (\ref{eq:ro_nfw}) along the line of sight, giving
\begin{eqnarray}
\label{eq:nfw}
\Sigma_{\mathrm{ nfw}}(X) = \frac{2 r_s \delta_c \rho_c}{X^2-1} \left\{ \begin{array}{ll}
\left(1-\frac{2}{\sqrt{1-X^{2}}}{\rm arctanh}\sqrt{\frac{1-X}{1+X}}\hspace{0.15cm}
 \right) 
& \mbox{$\left(X < 1\right)$} \\ 
 & \\
\left(1-\frac{2}{\sqrt{X^{2}-1}}\arctan\sqrt{\frac{X-1}{1+X}}\hspace{0.15cm}
 \right)
& \mbox{$\left(X > 1\right)$} 
\end{array}
\right.
\end{eqnarray}
\noindent and the shear is given by
\begin{equation}
\gamma_{\mathrm{nfw}}(X)= \frac{2 r_{s}\delta_{c}\rho_{c}}{\Sigma_{\mathrm{crit}}} \left[\frac{2}{X^2} \ln\left(\frac{X}{2}\right) - \frac{1}{X^2-1}+ 2\left(\frac{2}{X^2} +\frac{1}{X^2-1}\right) f(X)\right] \nonumber
\label{eq:nfw_shear}
\end{equation}
\noindent with
\begin{eqnarray}
\nonumber
f(X)=\left\{ \begin{array}{ll}
\left. \frac{ {\rm arctanh}\sqrt{(1-X)/(1+X)}}{\sqrt{(1-X^2)}} \right. & \mbox{$\left(X < 1\right)$}\\
 & \\ 
\left.\frac{{\rm arctan} \sqrt{(X-1)/(1+X)}}{\sqrt{X^2-1}} \right. & \mbox{$\left(X > 1\right)$}
\end{array}
 \right.
\end{eqnarray}
where $X=R/r_s$ \cite{wright}.  These lensing relations
of the NFW have previously been studied and can also be found in
e.g. \cite{bartelmann,wright,golse}.

\subsection{S\'ersic profile}
The S\'ersic law for surface density profiles is given by
\begin{equation}
\label{eq:sersic}
\ln \left(\frac{\Sigma_{\mathrm{ser}}}{\Sigma_e}\right) = -b_n\left[\left(\frac{R}{R_e}\right)^{1/n}-1\right],
\end{equation}
where $\Sigma_{\mathrm{ser}}$ is the 2D density, $R$ is the 2D radius,
$n$ is the S\'ersic index, $b_n$ is a constant chosen such that $R_e$
is the radius containing one-half of the projected mass and $\Sigma_e$
is the density at $R_e$.  The constant $b_n$ is found by solving the
equation $\Gamma(2n,b_n)=\Gamma(2n)/2$ and can be approximated as $b_n
\approx 2n - 1/3+4/(405n)+46/(25515n^2)$ \cite{ciotti}.  With $n=4$
the S\'ersic profile reduces to the de Vaucouleurs profile, whereas
$n=1$ gives the exponential law.  A previous study of the lensing
properties of the S\'ersic profile focussed on $n=3,4,5$\cite{cardone}.
In \cite{merritt} the mass distribution of dwarf- and
galaxy-sized halos has a mean of $n\sim3.0$ but  $n\sim2.4$ for
cluster-sized halos with values as low as $n=2$.

\section[]{Lensing and the S\'ersic profile}
We begin by calculating the total magnification, $\mu$, for the S\'ersic profile.  From standard equations for gravitational lensing \cite{schneider} we have:
\begin{equation}
\label{eq:mu}
\mu=\frac{1}{(1-\kappa)^2-\gamma^2}
\end{equation}
where $\kappa=\Sigma/\Sigma_{\mathrm{crit}}$ is the convergence and
\begin{equation}
\Sigma_{\mathrm{crit}}=\frac{{\mathrm c}^2}{4 \pi G}\frac{ D_s}{D_l D_{ls}}
\end{equation} 
is the critical mass density, which depends on the angular diameter
distances to the source ($D_s$), the lens ($D_l$) and between the
source and the lens ($D_{ls}$), $c$ is the speed of light in
vacuum and $G$ is the gravitational constant.  For a spherically symmetric mass profile, the shear $\gamma$ can be calculated as
\begin{equation}
\gamma=\frac{\bar{\Sigma}-\Sigma}{\Sigma_{\mathrm{crit}}}\equiv\bar{\kappa}-\kappa,
\end{equation} 
where
\begin{equation}
\bar{\Sigma}(Y)=\frac{2}{Y^2} \int_0^Y Y'\Sigma(Y') dY'
\end{equation} 
is the mean surface mass density as a function of radius  \cite{schneider}.

Applying this to the S\'ersic profile, we find
\begin{equation}
\label{eq:muser}
\mu_{\mathrm{ser}} = [(1-\bar{\kappa}_{\mathrm{ser}})(1+\bar{\kappa}_{\mathrm{ser}}-2\kappa_{\mathrm{ser}})]^{-1},
\end{equation}
where
\begin{equation}
\label{eq:kappaser}
\kappa_{\mathrm{ser}}=\frac{\Sigma_e}{\Sigma_{\mathrm{crit}}}\exp\left[-b_n\left(-1+Y^{1/n}\right)\right],
\end{equation}
\begin{equation}
\label{eq:gammaser}
\bar{\kappa}_{\mathrm{ser}}=\frac{\Sigma_e}{\Sigma_{\mathrm{crit}}}2 b_n^{-2n} n \exp[b_n]\frac{\Gamma[2n]-\Gamma[2n, b_n Y^{1/n}]}{Y^2}
\end{equation}
and $Y=R/R_e$.  We see from equation (\ref{eq:muser}) that the S\'ersic profile has two sets of critical curves (i.e. curves in the lens plane where the magnification formally goes to infinity), which are defined by $1-\bar{\kappa}_{\mathrm{ser}}=0$ (tangential critical curve) and $1+\bar{\kappa}_{\mathrm{ser}}-2\kappa_{\mathrm{ser}}=0$ (radial critical curve). 

 These equations define two rings and the radii can be found by
 solving the equations numerically.  The larger of the rings, given by
 $1-\bar{\kappa}_{\mathrm{ser}} = 0$, is the so called Einstein ring.
 Its radius is called the Einstein radius, $R_{\mathrm{ein}}$, which,
 for the spherical profiles considered here, corresponds to the radius
 at which the mean surface density equals the critical surface
 density.  
\begin{figure}  
\begin{center}
\epsfig{file=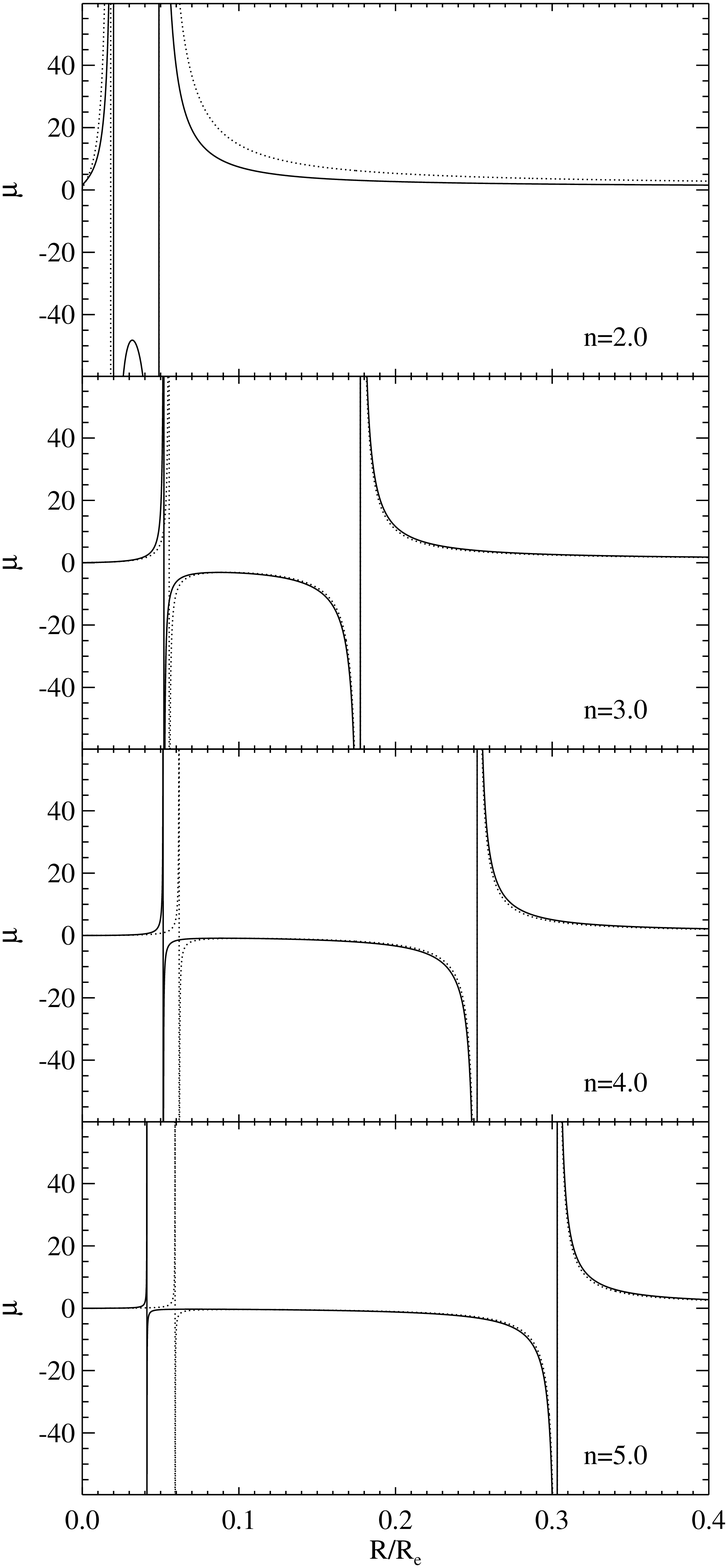, width=9.5cm}
\caption{Plot of the magnification properties of the S\'ersic profile (solid line) and the best fitting NFW profile (dotted line) for different values of the S\'ersic index $n$, and constant $\Sigma_{e}=10^{8}$M$_{\sun}$kpc$^{-2}$ and $R_{e}=100$~kpc.  We see that for the S\'ersic profile, the two critical curves move further apart for higher values of $n$, corresponding to a steeper profile.}
 \label{fig:muser}
\end{center}  
\end{figure}
Figure~\ref{fig:muser} (solid lines) shows a plot of the magnification as a function of radius for different values of n and constant $\Sigma_e/\Sigma_{\mathrm{crit}}$ and $R_e$.  We can see that the two critical curves move further apart for steeper profiles, and correspondingly, that the lowest magnification between them decreases.  
Our result is not in agreement with those of \cite{cardone}, who only
found tangential critical curves when studying lensing properties of
 the S\'ersic profile for $n=3,4,5$. We find that
this discrepancy is due to a sign error in the calculations of
\cite{cardone} and we note that, provided there is a
tangential critical curve at $R=R_{\mathrm{ein}} > 0$, there will always
be a radial critical curve for the S\'ersic profile.  This is because
$(1+ \bar{\kappa}_{\mathrm{ser}}-2 \kappa_{\mathrm{ser}})|_{R=\infty}
= 1 > 0$ and  $(1+ \bar{\kappa}_{\mathrm{ser}}-2
\kappa_{\mathrm{ser}})|_{R=0} = 1 - \kappa_{\mathrm{ser}}|_{R=0} < 0$ and as
both $\kappa_{\mathrm{ser}}$ and $\bar{\kappa}_{\mathrm{ser}}$ are continuous functions, there must exist an $R>0$ where a radial critical curve will occur.

\section{Comparing the S\'ersic and the NFW profiles}
In this section we compare the lensing properties of the S\'ersic and the NFW profiles.  We use the S\'ersic 2D mass profile as an input, and find a corresponding NFW which gives the best reproduction of the lensing constraints, as described below. 
Following \cite{graham}, values of  $R_e\sim100$~kpc with $3 \la n \la 5$ are representative parameter values for galaxy sized halos, while $R_e \sim 1000$~kpc with $2\la n \la 3.5$ are a realistic parameter range for the S\'ersic profile for cluster-sized halos, with lower $n$ being found for more massive clusters (which are more likely to act as lenses).  In particular, we take $500 \le R_e \le 2500$~kpc for cluster sized halos as input parameters for our simulations.  The third input parameter, $\Sigma_{e}$, has a minimum value for strong lensing to occur, i.e. $\Sigma_{\mathrm{ser}}|_{R=0} =\Sigma_{e} e^{b_n} > \Sigma_{\mathrm{crit}}$.  Typical Einstein radii for galaxies are of the order of $~1\arcsec$ \cite{lehar}, while for clusters they are of the order of $~10\arcsec$  with the largest known Einstein radius to date being $45\arcsec$  \cite{kneib, broadhurst}.  We will therefore constrain our input S\'ersic profiles to have Einstein radii in this range, and for each given Einstein radii we will vary $n$ and $R_{e}$, while $\Sigma_{e}$ is calculated from the three input parameters $R_{\mathrm{ein}}, n$ and $R_{e}$.
\begin{figure*}
\epsfig{file=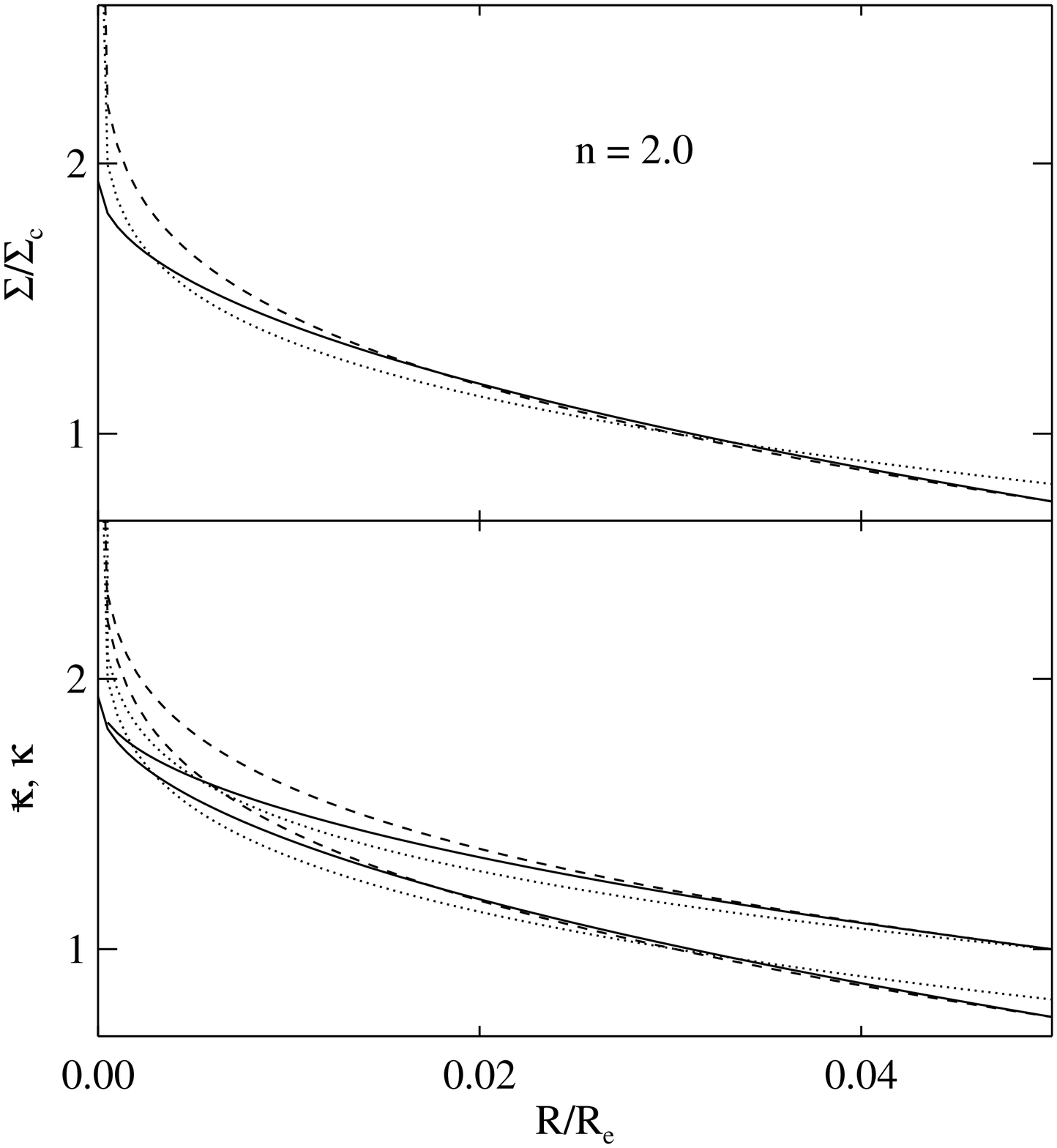, width=8.2cm}
\epsfig{file=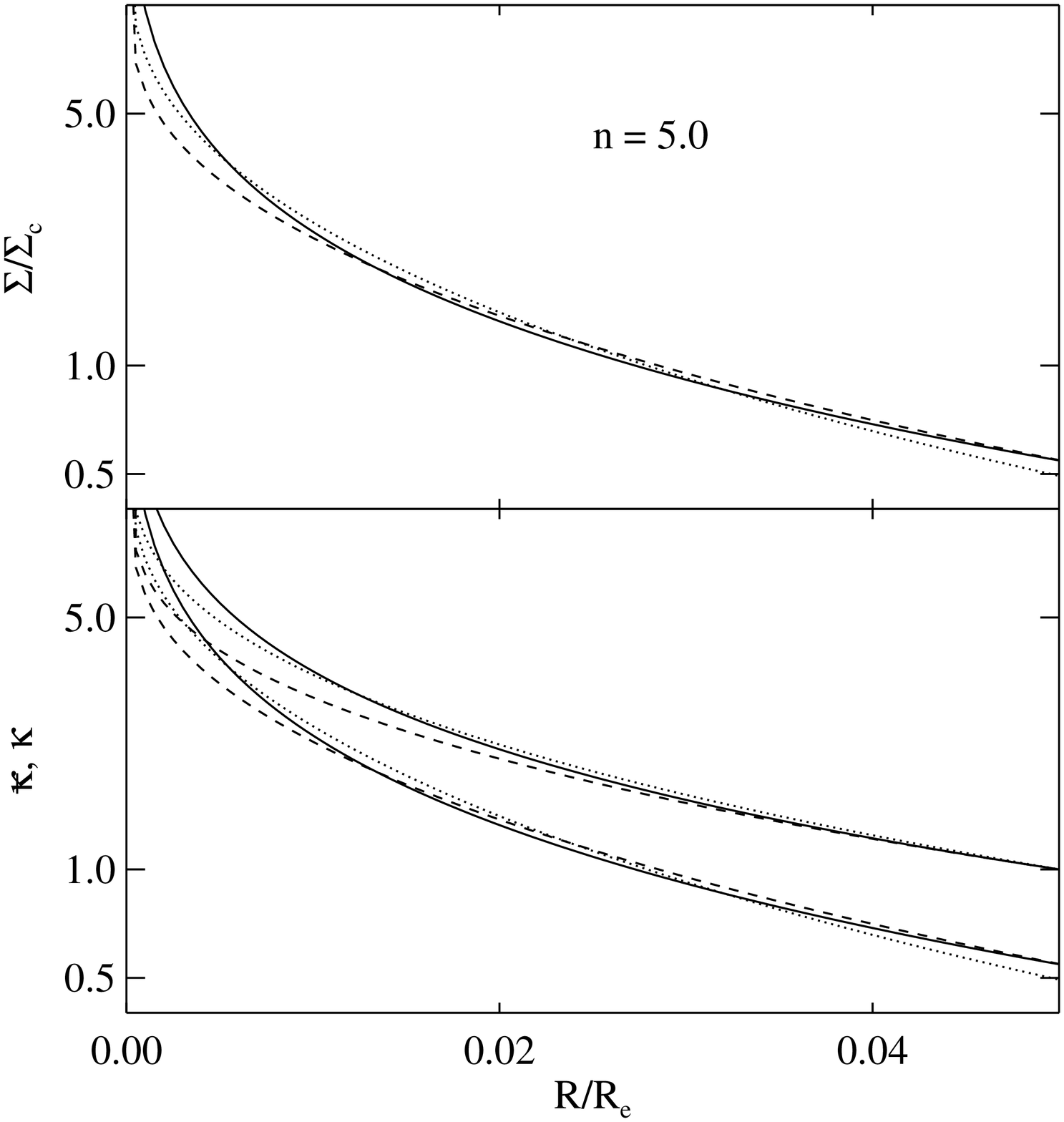, width=8.2cm}
\caption{The top panels show $\Sigma$ for the S\'ersic profile (solid line) for n=2,5 and $0\le
  R\le R_{\mathrm{ein}}$.  Also shown are the best fitting NFW profile (dotted line) as defined in Section \ref{sec:fitting} and an alternative fit (dashed line) which is optimised in a tight area around $R_{\mathrm{ein}}$.
  The lower panels show the corresponding $\bar{\kappa}$ (upper set of curves)
  and $\kappa$ (lower set of curves).  Although the detailed shape of the NFW varies with the choice of fitting, the resulting analysis presented in the paper is not sensitive to it.  The value of the other input parameters for the S\'ersic profile were $R_e=1000$~kpc and $R_{\mathrm{ein}}=50$~kpc.}
 \label{fig:sigma}
\end{figure*}

Our goal is to study whether strong lensing data of an accuracy
currently available can be well fitted by an NFW profile even if the
lensing mass distribution actually follows a S\'ersic profile.  In
particular, we wish to explore whether the corresponding NFW can
reproduce the strong lensing data, and if so, what the parameters
of the NFW profile are and how these relate to typical NFW parameter values obtained for simulated cluster halos.  Therefore we choose to take a S\'ersic profile as our assumed input and will fit for the corresponding NFW profile given certain constraints.

The constraints from strong lensing are in general the Einstein
radius, relative locations of the images and their magnification
ratio, which are all sensitive to the mass distribution within the
disc defined by the outermost image.  Of these the Einstein radius
(deduced from the total separation of the multiple images) is the best
constrained value, while the precise location of images can be
affected by nearby substructure and the magnification ratio can be
biased by both microlensing and dust extinction in the lensing galaxy
\cite{schechter,eliasdottir,lewis}. The inner critical
curve may also give additional constraints on the mass profile if
central images or merging radial arcs are observed.  Under the
assumption of spherical symmetry, time delay
measurements between images provide a further constraint on the mass
in an annulus, bounded by the radii of each of the images, but are
hard to measure, in particular in strong cluster lensing where most of
the known lensed images are background galaxies and not quasars.

We choose two main approaches to the fitting.  The first approach assumes that we have multiple systems at different redshifts allowing for constraints on the total mass within a number of radii.  This approach reflects the common situation that several arc systems are used to constrain the lensing mass distribution of clusters \cite{limousin}.  The second approach assumes that only a single Einstein ring is detected and used for the lensing analysis \cite{comerford2007}.  In this case we fix the Einstein radii of the two profiles to be the same, and fit the inner mass density profile as described below in section 3.1.  This choice is motivated mainly by the fact that the Einstein radius is easy to constrain accurately observationally and that the magnification of multiply imaged arcs and arclets in strong lensing clusters constrain the surface mass density on a scale given by the Einstein radius.

\subsection{Fitting method}
\label{sec:fitting}
For the case when only one Einstein ring is used to constrain the system, the fit is underconstrained, so we must choose an additional constraint on the NFW profile that relates it to the input S\'ersic profile. Since the magnification and shear for a series of observed extended and highly magnified arcs of the same source is sensitive to the surface mass density in the inner regions of the cluster, we require that the NFW profile fits the input S\'ersic projected mass distribution in an inner region bounded by $R_{\mathrm{ein}}$ with the Einstein radius itself as a constraint.  The fitting itself is done numerically, using  Levenberg-Marquardt
least-squares minimisation.  As our primary mode of fitting, we take
the S\'ersic surface mass density profile at
$N=10000$ points, linearly distributed from $R=0$ to $R=R_{\mathrm{ein}}$, and applying a constant relative error in each of the points.  To check the
dependence of the results on our precise choice of fitting, we try
several other fitting methods consistent with the previously described
framework.  In particular, we redo the analysis using fits where we
distribute the points differently, where we fit $\log \Sigma$ instead
of $\Sigma$, and where we alter the weight of the points, giving more
emphasis to different regions.  For example, instead of distributing
the points over $0<R<R_{\mathrm{ein}}$, we constrain the fit to a
tight area around the Einstein radius ($0.95 R_{\mathrm{ein}} < R <
1.05 R_{\mathrm{ein}}$) or to an area around the Einstein radius and
the inner critical curve of the S\'ersic profile.  In addition we
changed the absolute weights of the points to be equal, putting an
emphasis on the inner points (where we have less constraints from
lensing).  Although the detailed shape of the best fitting NFW profile
 varies depending on the choice of
fitting (see Figure \ref{fig:sigma} for a plot of the $\Sigma, \bar{\kappa}, \kappa$ from two fits for $n=2$ and $5$), and we note that  we do not constrain whether the profile is very cuspy or not in the very centre as could be done using central images as constraint, the
analysis presented in the rest of the paper remains qualitatively the same.  This leads us to conclude
that the precise method of fitting we use does not affect our
conclusions. 

When we have multiple Einstein radii, we have enough constraints to find the best fitting NFW without additional constraints.  As each Einstein radii gives an estimate of the mass of the system within that radius, we fit the mass profile (rather than the mass density
profile) of the NFW to the mass profile of the input S\'ersic profile.
We assume that we have 9 independent systems at distances of
$10$-$170$~kpc for a cluster sized lens \cite{limousin}.
We set the uncertainty of the mass estimated at each of the points to
be $10\%$, corresponding roughly to the level of current observational
accuracy. We find that the results from the lensing analysis
remain qualitatively the same as for our fitting of a single Einstein radius.

 The figures in the paper are based on the single Einstein radius constraint, fitting $\Sigma$ in the range of $0<R<R_{\mathrm{ein}}$ using the $1/N$ weighting,
unless otherwise noted.

\subsection{General properties}
\begin{figure}  
\begin{center}
\epsfig{file=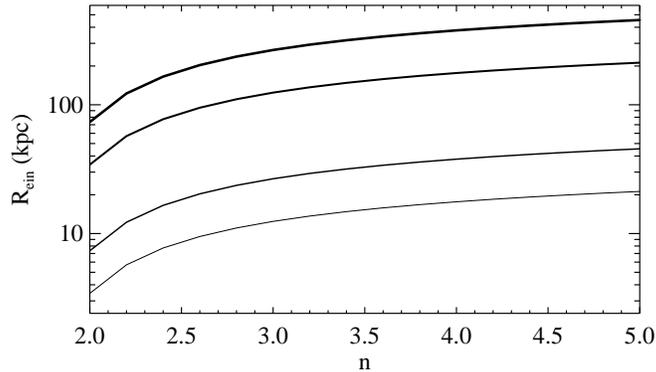, width=9.5cm}
\caption{ The Einstein radius, $R_{\mathrm{ein}}$, as a function of $n$, for $R_{e} = 70, 150, 700, 1500$~kpc (denoted by increasing line thickness) and constant $\Sigma_{e}=10^{8}$~M$_{\sun}$~kpc$^{-2}$.  The Einstein radius is strongly dependent on $R_e$, with larger values of $R_{e}$ resulting in a larger Einstein radius, and is also dependent on $n$, with larger $n$ giving larger $R_{\mathrm{ein}}$.  The $n$ dependence is strongest for low $n\sim 2$--$3$ but levels off for higher $n$.  A similar plot can be obtained by
 holding $R_{e}$ constant and varying $\Sigma_{e}$, showing $R_{\mathrm{ein}}$ increasing with $\Sigma_{e}$.}
 \label{fig:Rein}
\end{center}  
\end{figure}
Figure~\ref{fig:Rein} shows $R_{\mathrm{ein}}$ as function of $n$ for different values of $R_{e}$.  The figure shows that $R_{\mathrm{ein}}$ is a strong function of $R_e$ and $n$ for low values of $n$, while for the higher range of $n$ the values level off.  A similar plot may be obtained keeping $R_{e}$ constant and varying $\Sigma_{e}$, which shows $R_{\mathrm{ein}}$ increasing with $\Sigma_{e}$.
As explained above, our interest in the strong lensing properties of
the S\'ersic profile as compared to the NFW profile motivates us to
either fix the Einstein radius or fit for the total mass at multiple
Einstein radii when making further comparisons. In the following
figures the Einstein radius is set to 
$R_{\mathrm{ein}}=50\,\mathrm{kpc}$, corresponding to a cluster sized
halo, but all the results presented also apply to systems with different
Einstein radii (e.g. galaxies) and to systems with measurements of
several Einstein radii corresponding to multiple sources.

Fitting an NFW profile to a S\'ersic lens will not always give a good
fit to the data. In fact, given that the number of parameters of the
NFW is less than that of the S\'ersic one does not expect a good fit
in general. Different indices $n$ of the S\'ersic profile correspond
to different slopes in the inner mass profile, whereas the slope of
the NFW profile is not a free parameter.
\begin{figure}  
\begin{center}
\epsfig{file=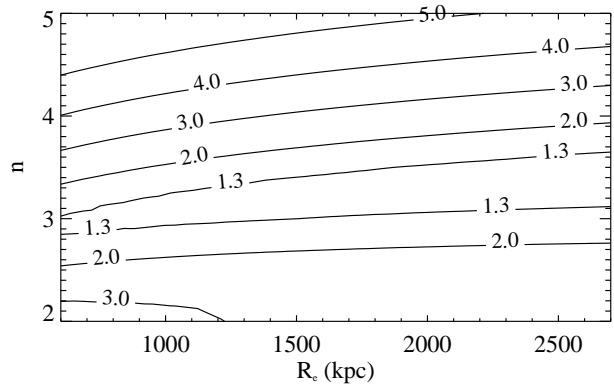, width=9.5cm}
\caption{The reduced $\chi^2$ of the best-fit NFW as a function of the S\'ersic parameters $n$ and $R_{e}$, normalised with the lowest $\chi^2$ in the ensemble.  Some set of parameters give significantly worse fits than others, consistent with the fact that the number of free parameters for the NFW is less than for the S\'ersic profile.   The Einstein radius is kept fixed at
$R_{\mathrm{ein}}=50\,\mathrm{kpc}$, as described in the text.}
 \label{fig:chi}
\end{center}  
\end{figure}
 In Fig.\,\ref{fig:chi} we show the  'goodness of fit', the $\chi^2$ normalised with the lowest $\chi^2$ in the ensemble.
As expected, good fits (low values of $\chi^2$)
are obtained only for a small region around a specific value of $n$.
We note, however, that in most lensing analysis an NFW profile is
assumed and the best-fit parameters are obtained, even if the actual
$\chi^2$ is high. In such cases, our
results suggest that a fit using a S\'ersic profile should be used
instead and will give a much better fit if the underlying profile is
in fact a S\'ersic.
\begin{figure}  
\begin{center}
\epsfig{file=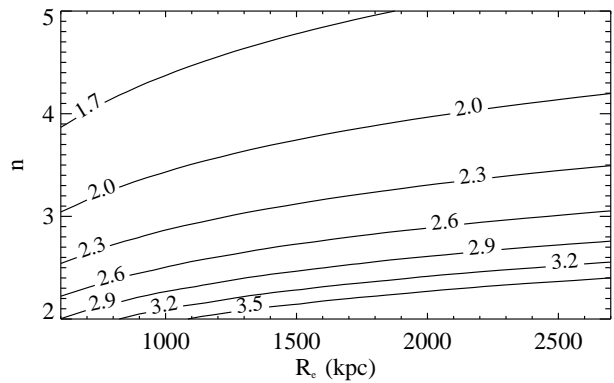, width=9.5cm}
\caption{Scale lengths, $\log \left(r_{\mathrm{s}}/\mathrm{kpc}\right)$, for the best-fit NFW profile
as a function of S\'ersic parameters, $R_e$ and $n$.  The scale length is strongly dependent on $n$, with low $n$ giving rise to high scale lengths.  The Einstein radius is fixed
as in the previous figures.}
 \label{fig:rs_comp}
\end{center}  
\end{figure}
Figure~\ref{fig:rs_comp} shows how the best-fit NFW scale
length varies with the input S\'ersic parameters. Higher values of $n$
give rise to steeper profiles and hence a lower value for the scale
length $r_s$ of the best-fit NFW profile. 

\subsection{Magnification and image configuration}

In Figure~\ref{fig:muser} we overplot the magnification for the best fit NFW profile for $n=2,3,4,5$ and constant $R_{e}$ and $\Sigma_{e}$ (note that for this plot we do not keep $R_{\mathrm{ein}}$ constant).  The figure shows that the difference in
the magnification properties is not very large, except for $n=2$.  For
$n=3$, the fits are very similar, whereas for $n=4,5$ the main
difference lies in the location of the inner critical lines.  The
exact  location of the inner critical lines depends to some extent
also on our choice of fitting, and is therefore not a very robust
measurement of differences in the two profiles. In addition, the inner
critical line is usually observationally less well determined than the
Einstein radius. Measurements of different Einstein radii, due to
sources at different redshifts, are a more common and practical
additional constraint on the inner mass profile. We found that constraining the
mass profiles in this way at several different Einstein radii results in very
similar magnification profiles.

\begin{figure}  
\begin{center}
 \epsfig{file=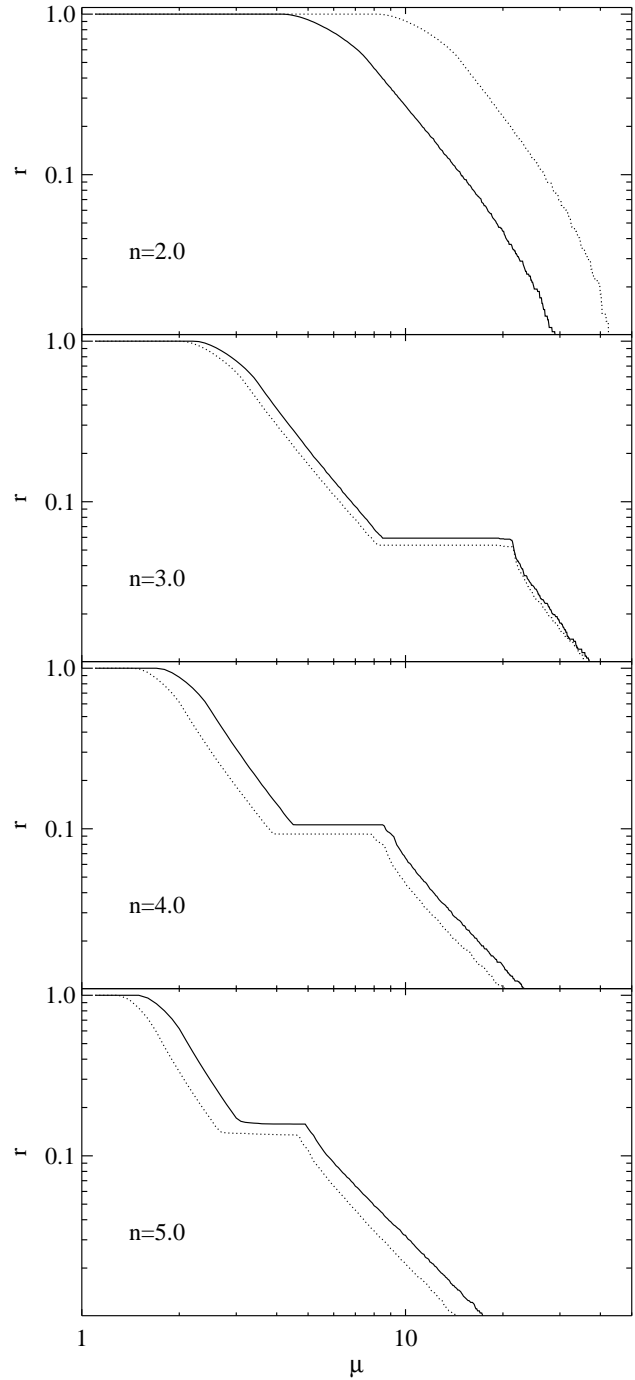, width=9.5cm}
\caption{The ratio of the number of images with amplification equal or greater than $\mu$ to the total number of amplified images,   
 $r(\mu)$, for the S\'ersic (solid line) and the NFW (dotted line) profiles for different values of $n$.  The calculation of $r(\mu)$ is done in the source plane, corresponding to magnification maps of size $4 R_{\mathrm{ein}}$ times $4 R_{\mathrm{ein}}$ in the image plane, shown in Figure~\ref{fig:plot_images}, with $R_{\mathrm{ein}}=50$~kpc and $R_{e}=1000$~kpc (corresponding to a cluster sized lensing halo).  The plots show that the magnification properties are similar for the two profiles, although the NFW over predicts highly magnified images for $n\sim2$, while it to a smaller degree under-predicts them for $n\ga3$.}
 \label{fig:mag}
\end{center}  
\end{figure}
\begin{figure*}
\epsfig{file=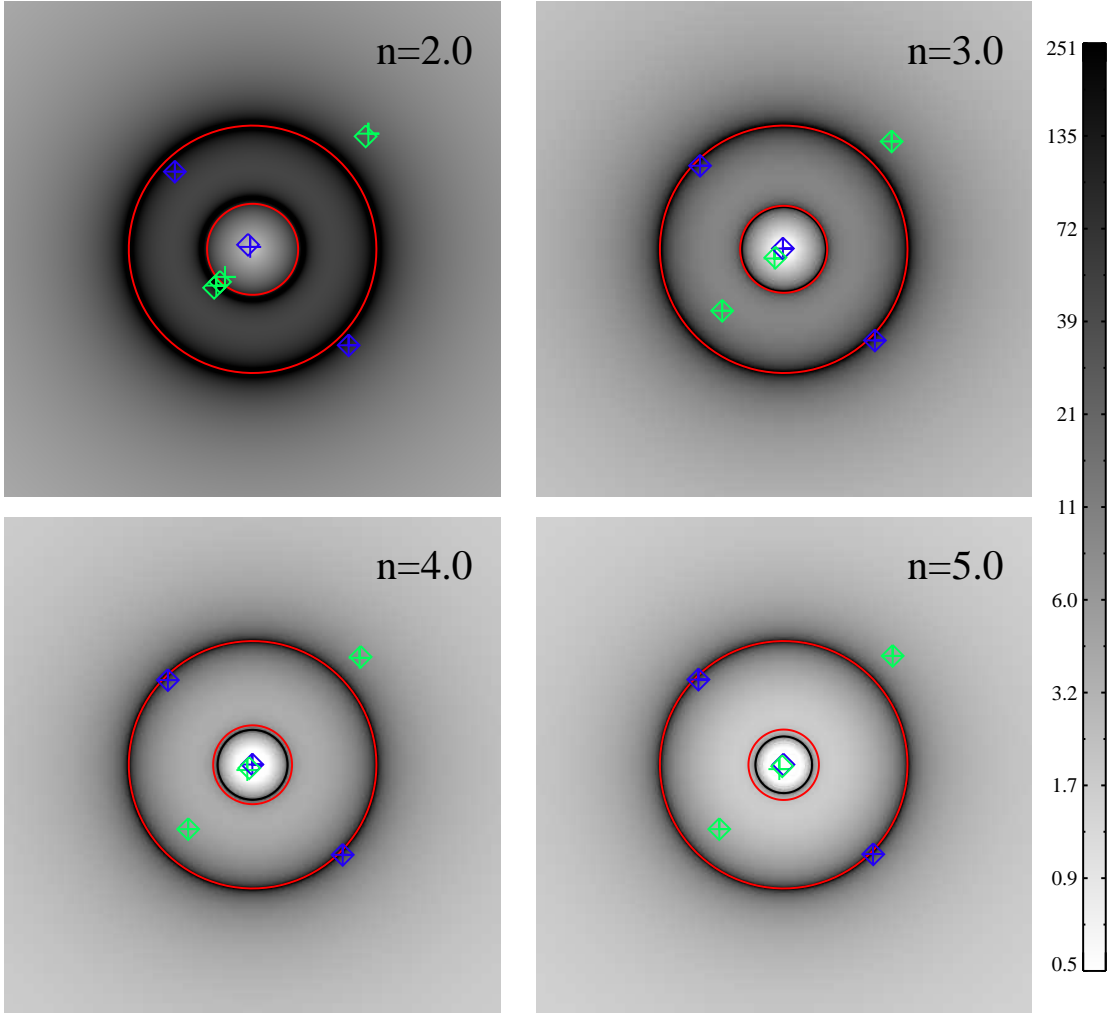, width=15cm}
\caption{The image plane magnification map
for the S\'ersic profile for different values of $n$ with $R_{\mathrm{ein}}=50$~kpc and $R_{e}=1000$~kpc (corresponding to a cluster sized lensing halo).  The images are centred on the lens and are of size $4R_{\mathrm{ein}}$ times $4R_{\mathrm{ein}}$.  The black circles correspond to the tangential and radial critical curves for the S\'ersic profile, and the red circles are the critical curves of the best fit NFW profile, where the magnification formally goes to infinity.  Also shown are the image geometries for double images of the S\'ersic (diamonds) and the NFW (plus signs) profiles for two different source positions (blue and green).  The symmetrical configuration of images are not distinguishable for any $n$, whereas for an asymmetric configuration the images are only marginally distinguishable for $n=2$.}
 \label{fig:plot_images}
\end{figure*}

For a more quantitative analysis of the magnification differences, we
define $r(\mu)$ as the ratio the area on the source plane where magnified images (i.e. images with $\mu\ge1$) with magnification
greater or equal to $\mu$ occur.  Taking $R_{\mathrm{ein}}$ to be constant, in Figure~\ref{fig:mag} we plot $r(\mu)$ calculated over an area in the source plane, corresponding to $4 R_{\mathrm{ein}}$ times $4 R_{\mathrm{ein}}$ in the image plane (shown for the S\'ersic profile in Figure~\ref{fig:plot_images}).  Figure~\ref{fig:mag} shows that the difference in the magnification is indeed small, although for $n=2$ the NFW profile is more likely to produce highly magnified images, whereas for higher $n$ the reverse is true to a smaller degree. 
As the difference in the magnification properties of the S\'ersic and
the best fitting NFW profiles is small, our results demonstrate that,
given current observational accuracy, an NFW profile is adequate to describe the magnification for
these strong lensing systems. 

In many strong gravitational lens systems the observed images do not
lie exactly on the Einstein radius, but are displaced in an
asymmetric fashion; one image being within the Einstein radius and
another outside. To illustrate the difference between S\'ersic and NFW
models we show the image displacements in 
Fig.\,\ref{fig:plot_images} for
two different choices of the source position. For a fixed source
position, the image positions for the best-fitting NFW profile is
indistinguishable from that of the S\'ersic profile for high $n$, while for low values of $n$, there is a slight offset between the
image positions, in line with the findings from
 above:
NFW profiles produce very similar lensing signatures to S\'ersic
profiles, and are distinguishable only for low values of the S\'ersic
index $n\la3$ unless the goodness of fit can be used to separate the profiles.

\subsection{Mass and concentration}
Lensing, both strong and weak, is frequently used to estimate the mass and the NFW concentration parameter of clusters, which are then compared to those from simulations or X-ray measurements, with lensing usually finding a higher mass and concentration \cite{kneib,kling2005, halkola2006,limousin, comerford2007}.  Therefore, it is of interest to study whether an underlying S\'ersic profile could cause a bias in the mass and $c$ estimate, if an NFW profile is assumed in the strong lensing modelling.  As strong lensing is only sensitive at small radii, one should keep in mind when making these comparison, that a small deviation in the assumed profile can lead to large errors in both the mass and concentration parameter.  However, although weak lensing should be more accurate for such comparisons as it probes a similar distance scale, it is very sensitive to contamination of the background galaxy catalogue by cluster members.  In fact, \cite{limousin} find for Abell 1689, that the concentration parameter derived from strong lensing is consistent between different data sets and methods, while that derived from weak lensing still shows large scatter which are an artefact of the method applied.
\begin{figure}  
\begin{center}
\epsfig{file=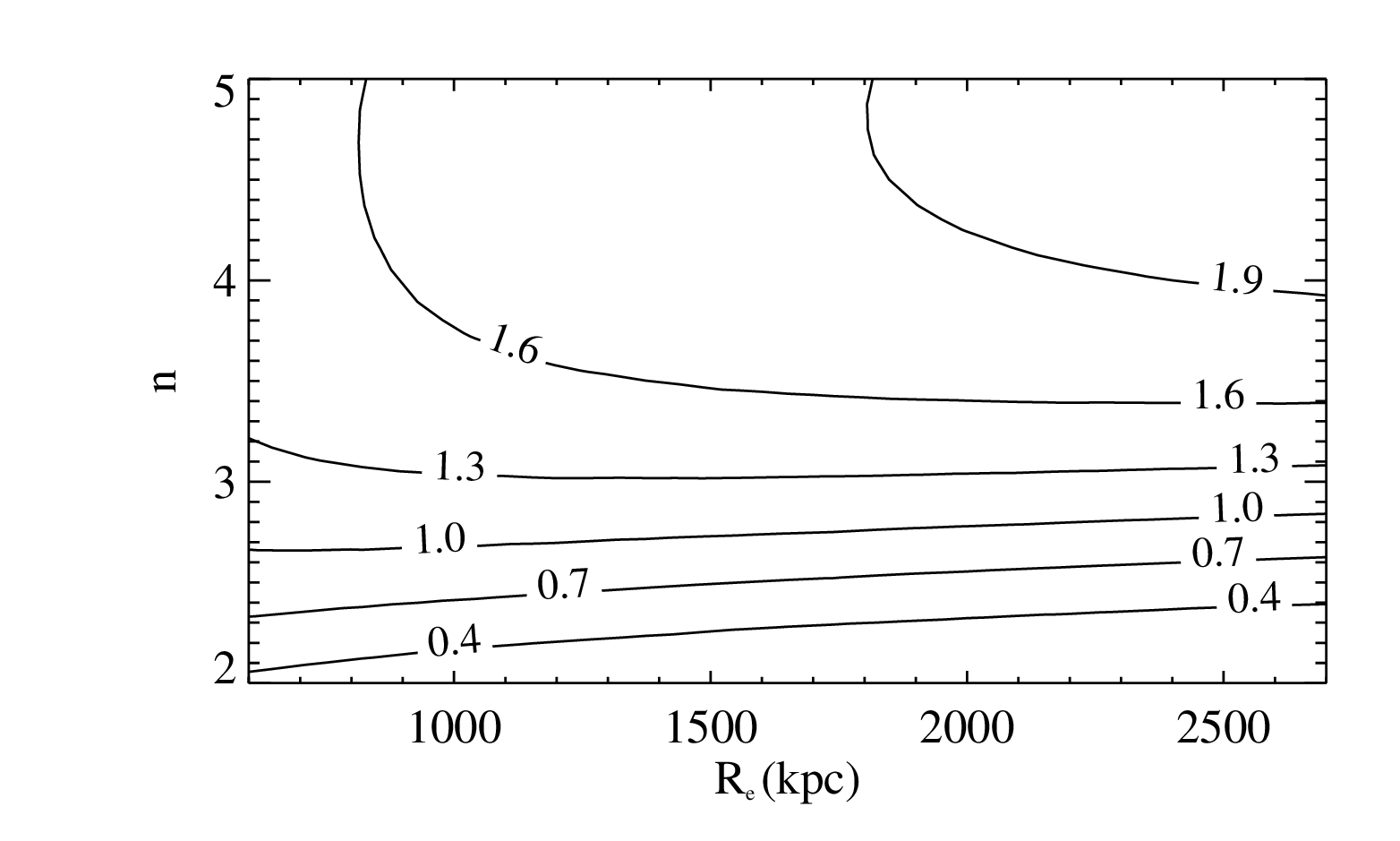,  width=9.5cm}
\caption{The ratio of projected S\'ersic input masses over the best-fit NFW
masses within $r_{200}$ as a function of $R_e$ and $n$ of the input S\'ersic profile.   For low $n$ the NFW profile overestimates the mass by a factor of around $\sim 2$, while the opposite holds for high $n$.  The Einstein radius is kept fixed at
$R_{\mathrm{ein}}=50\,\mathrm{kpc}$, as in the previous figure.  }
 \label{fig:mass_comp}
\end{center}  
\end{figure}
In Fig.\,\ref{fig:mass_comp} we show the ratio of the projected masses
for the input S\'ersic and the best-fit NFW within $r_{200}$ (of the NFW profile). The mass ratio depends
most strongly on the S\'ersic index $n$. Roughly speaking, a higher
value of $n$ gives rise to lower projected masses for the
corresponding NFW fits.  Therefore, applying an NFW profile to a
S\'ersic lens can lead to a significant error in the mass estimation,
with the mass at $r_{200}$ being overestimated by a factor of $\sim 2$ for low $n$.
\begin{figure}  
\begin{center}
\epsfig{file=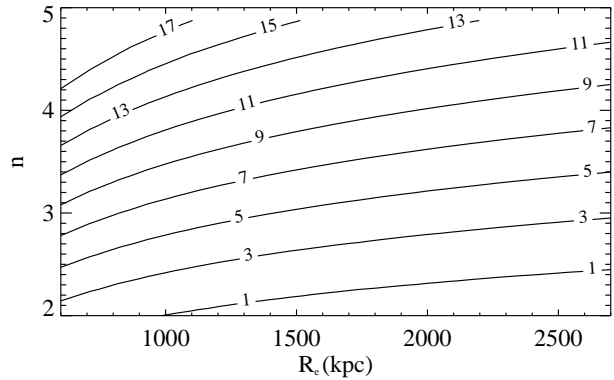, width=9.5cm}
\caption{Concentration parameter for the best-fit NFW halo as a
  function of the input S\'ersic profile parameters $n$ and $R_e$.  For low $n$, the concentration parameter takes on values of $c$ which are lower than those expected from simulations, while for higher $n$ the values of $c$ become higher than expected.  Therefore, although the lensing properties can be reproduced, the parameters of an NFW fit to a S\'ersic lens can become unrealistic (see also Figs.~\ref{fig:mass_comp} and \ref{fig:rs_comp}).
  The Einstein radius is kept fixed as in the previous figure.}
 \label{fig:c_comp}
\end{center}  
\end{figure}
Similarly, Figure~\ref{fig:c_comp} shows the concentration parameter of the best
  fit NFW profile as a function of the S\'ersic index $n$ and scale
  $R_e$. The figure shows how the concentration parameters increase
  mostly with increasing $n$ from very low concentrations to $c\sim17$
  for high $n$.    Extreme values of the concentration parameter (i.e. $c<3$ or $c>10$), can be found and correspond to the input S\'ersic parameters where the NFW fit is worse and where the mass estimate is biased (see Figs.~\ref{fig:chi} and \ref{fig:mass_comp}).
  
 Therefore applying an NFW profile
  to an underlying S\'ersic profile, can lead to wrong mass estimates.
  For cluster sized halos with $n\sim2$, fitting lensing data with an
  NFW profile would give values of $c$ which would be lower than those expected when compared to CDM simulations.   Such low values of $c$ have been found in strong lensing analysis, although in general lensing overpredicts $c$ compared to X-ray measurements \cite{hennawi, comerford2007}. 
For higher values of $n$, this trend gets reversed:
  for $n\ga3$ concentration parameters of NFW fits are higher than those seen
  in simulations.   However, such high values of $n$ are not expected to describe cluster size halos, but galaxy scaled halos \cite{merritt}.

Observationally it is thus likely that NFW lens models of
  clusters that give rise to mass or concentrations which
  seem inconsistent or only marginally consistent with
  simulations or X-ray measurements would  be better described by a S\'ersic
  profile. We stress that these results are consistent for all the fitting methods we apply, and whether we apply one or several sets of Einstein radii as constraints.

\subsection{Extrapolating to the weak lensing regime}
\label{sec:weak}
\begin{figure}  
\begin{center}
\epsfig{file=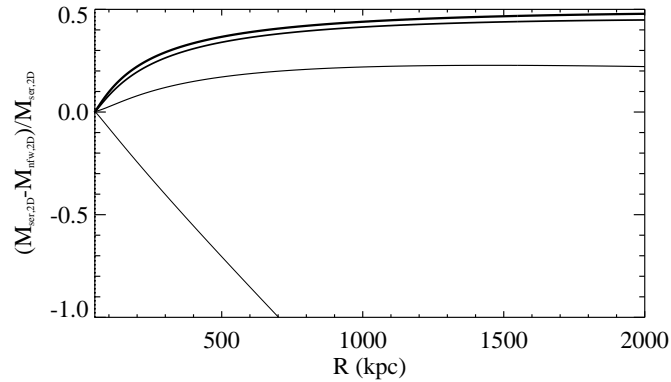, width=9.5cm}
\caption{The difference in the projected mass vs. the mass of the S\'ersic profile at radii $R>R_{\mathrm{ein}}$ for $n=2,3,4,5$ (denoted by increasing line thickness) with $R_{\mathrm{ein}}=50$~kpc and $R_{e}=1000$~kpc.  The NFW fit to the S\'ersic profile, extrapolated to higher radii, underestimates the mass for an underlying S\'ersic profile for $n\ga3$, whereas it overestimates it for low $n$.}
\label{fig:high_R_mass}
\end{center}  
\end{figure}

We have  focused on comparing the difference in the strong lensing properties of the S\'ersic and the NFW profiles, making $R_{\mathrm{ein}}$ the most relevant distance scale.  A thorough analysis of the difference in the weak lensing properties of the two profiles would require optimising the fits in the weak lensing regime and is beyond the scope of this paper.  However, we can compare the predicted total mass and the weak lensing shear $\gamma$ in the weak lensing regime by extrapolating the derived functions to higher radii.  We have already compared the projected mass within $r_{200}$ for different values of $n$ and $R_{e}$ (see Figure~\ref{fig:mass_comp}), but here we study the NFW: S\'ersic mass ratio as a function of radius for a halo with $R_{e}=1000$~kpc and $R_{\mathrm{ein}}=50$~kpc, corresponding to a cluster sized halo.
In Figure~\ref{fig:high_R_mass},
 we plot the difference in the projected mass over the projected mass of the S\'ersic profile.  We see that for $n=2$, the best fitting NFW profile overestimates the mass by a factor of $> 2$ for $R \ga 10R_{\mathrm{ein}}$, while it underestimates it by $\sim 2$ for $n\ge3$.  This is due to the fact that in order to fit the low $n=2$ S\'ersic profiles, the estimated $r_{s}$ of the NFW (the scale radius at which the density slope increases) is overestimated, making the NFW profile fall off less steeply than expected in the outer regions.  Similarly, 
\begin{figure}  
\begin{center}
 \epsfig{file=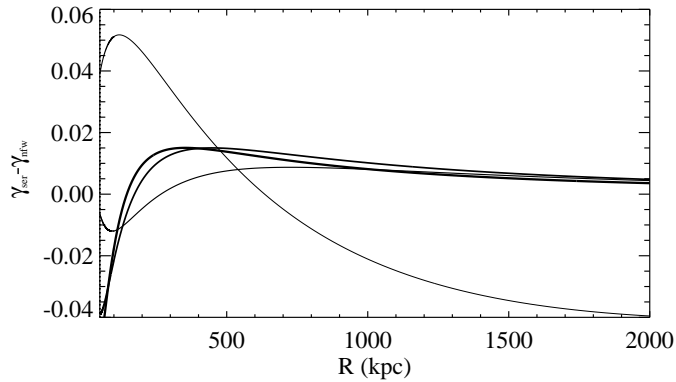, width=9.5cm} 
\caption{The difference in the shear of the S\'ersic and NFW profiles at radii $R>R_{\mathrm{ein}}$ for $n=2,3,4,5$ (denoted by increasing line thickness) with $R_{\mathrm{ein}}=50$~kpc and $R_{e}=1000$~kpc.  For $n\ga3$ the NFW fit to the S\'ersic profile, extrapolated to higher radii, underestimates the shear by $\sim 0.01$.  For low $n$, the shear can be more strongly affected: it is underestimated by up to  $\sim0.05$ for low radii and overestimated by a similar amount for $R\ga10 R_{\mathrm{ein}}$.  }
 \label{fig:high_R_gamma}
\end{center}  
\end{figure}
in Figure~\ref{fig:high_R_gamma}, we plot the difference in the
projected shear $\gamma$ as a function of radius.  We find that for
$n\ga3$, the difference is of the order of $0.01$ while for $n=2$ it
is of the order of $0.05$.  As weak lensing measurements of clusters
are usually looking at shear signals around $0.01-0.05$, this is a
significant difference, in particular for the low $n$.  Therefore,
using an NFW to model an underlying S\'ersic profile with low $n$
would significantly over predict the weak lensing signal at large
radii, and, depending on the measurement errors, result in a poor fit.

\subsection{Observational implications}
\label{sec:obs}
We have taken a S\'ersic profile as the underlying description of the matter profile and studied how lensing analysis applying an NFW fit would differ and lead to biased results, depending on the input S\'ersic index.  In particular we have found that for $n\sim2$ the mass can be overestimated by a factor of two and give low concentration.  We have however not addressed which profile is a better description of real dark matter halos.  To do so, one should model a real system using the S\'ersic profile and compare it to the results found when using an NFW profile.   While such a study is beyond the scope of this paper, we discuss here the implications our results have on lens modelling of real systems.

For galaxy scale lenses, various authors \cite{chae2002,li2002,koopmans2006} have established that lenses on average are well fitted by an isothermal profile, although there can be significant scatter for individual lenses \cite{koopmans2006}.  These all apply however to the {\it total} matter distribution, both luminous and dark, while we have been studying the differences for dark matter only profiles.  More recent studies have attempted separating the dark and luminous matter profiles in lensing of galaxies \cite{baltz2007,dye2007}, but in general the constraints from lensing alone are not sufficient for separating the two.

Group or cluster scale lenses, with which we are primarily concerned in this work, are more dark matter dominated. This means they are also better suited for an observational test of whether real clusters are better described by an NFW or a S\'ersic profile.  Regarding observational strong lensing constraints the main observable is the total mass and the slope of the mass profile in the innermost (strong lensing) region of the cluster or group. The NFW and the S\'ersic profile differ in that the S\'ersic profile has an extra independent parameter, $n$ that directly relates to the inner slope of the mass distribution. Even though the concentration parameter of the NFW profile is related to the slope of the profile, it is also related to the overall scale and mass of the cluster and is not an independent parameter. From an observational perspective the key quantities then are the slope and mass of the dark matter in the inner region of real clusters, and whether they are consistent with an NFW profile as expected from CDM simulations. As pointed out recently \cite{comerford2007} NFW fits to observed strong lensing clusters very often give masses or concentration parameters that seem inconsistent with CDM predictions. In general the trend is that observed clusters have higher concentration parameters than predicted in simulations -- but some lensing clusters have also yielded very low concentration parameters when fit with an NFW profile (e.g. MS1137.5+6625 \cite{maughan2006} and MS 2053.7-0449 \cite{verdugo2007}).
Clusters with unusually high or low concentration values may very well be fit much better with a S\'ersic profile -- since the Einstein radius then becomes a function of $n$, even for a fixed scale, $R_e$, an index $n$ (and hence slope) and a total mass $M$ may be found that match the predicted shape of the mass profile from CDM more closely.   It is important to note that the lensing fits of clusters are often quite poor, i.e. the $\chi^2$ is high or the rms of the predicted image position is larger than the uncertainty in their measured positions \cite{limousin, verdugo2007}.

\section{Discussion and Conclusions}
\label{sec:summary}
In this paper, we have compared the S\'ersic and NFW dark matter surface density
profiles with respect to their strong lensing properties. Taking S\'ersic profiles, with parameters that are in the range found for clusters and galaxies in large N-body simulations, we explored the parameters of the NFW profile that fits the inner projected surface mass density of the S\'ersic best. The NFW profiles were constrained to have the same Einstein radius as the input S\'ersic profile, as this is the most accurately determined observable in lens systems.  
As an alternative approach, we have assumed a data set consisting of several multiply imaged systems at different redshifts, giving a constraint on the mass profile at different radii. 
We find that an NFW profile can in general accurately produce the
magnification and image positions of a S\'ersic matter distribution
only for $n\ga3$.  For lower S\'ersic index values, which are more
likely to occur in cluster sized halos, the difference in the lensing
properties increases, to the level where it could affect results of
lens modelling, in particular for the magnification estimates.  We also note that although one can in general find an
NFW profile which gives similar lensing properties, the parameters of
the profile may become unrealistic.  For $n\sim2$, as simulations suggest can be the case for clusters, the mass is overestimated by a factor of $\sim2$ while the concentration parameter is $\sim1$-$3$, which is lower than expected from simulations, although such low values have been seen in lensing studies \cite{comerford2007}.
 Extrapolating into the weak lensing regime we find that the NFW overestimates both the total mass and the shear for $n=2$, while it underestimates them for $n\ga3$.  
 
Therefore, if an underlying S\'ersic profile with $n\sim2$, as is found for massive systems, is fit with an NFW using strong lensing constraints, it will overpredict both the mass (by a factor of a few) and the shear (up to $\sim 4\%$) of the halo at large radii, and could therefore contribute to explaining why lensing mass estimates are greater than those found by X-rays.  Conversely, weak lensing data can in principle be used to demonstrate whether the S\'ersic or the NFW profile is a more accurate description of the mass distribution. 
However, in practice a good S/N ratio is required -- even for S\'ersic profiles with low n an accuracy of better than $\sim4\%$ is needed to distinguish the profiles and identify the shape of the mass profile accurately enough to obtain reliable parameters. Most weak lensing data is not yet of sufficient accuracy to rule out an NFW profile as a good representation of the mass distribution even if the mass distribution is in fact a S\'ersic profile with low $n$.   

In our analysis we have neglected the contribution of baryons -- our results are therefore most applicable to massive systems, like clusters of galaxies where the dark matter dominates. Even though our results clearly show that for a range of S\'ersic parameters, in particular for halos with low $n$ (corresponding to the most massive halos), strong lensing data can not always be well reproduced by an NFW profile, we do not address the question which of the two mass distribution is in fact a better fit to the mass distribution in clusters. However, the results from N-body simulations suggest \cite{merritt} that the S\'ersic profile is a better fit, at least in the inner regions, which are also the regions probed by strong lensing. Even if the overall mass profile of clusters is often fit well by an NFW profile, the NFW profile may not reproduce the mass distribution in the inner part adequately. As demonstrated here,  the sensitivity of the strong lensing properties on the form of the mass profile in that region then means that fitting an NFW profile to the lensing data will lead to rather meaningless parameters; the concentration parameters and masses of NFW profiles recovered in this way \emph{can not} simply be compared to the values obtained in simulations. In this work we have for the first time discussed the strong lensing properties of the S\'ersic profile, given important lensing relations, and compared it to the NFW profile. The differences are important enough to warrant the inclusion of  S\'ersic profile for future analysis of strong lensing clusters.

\begin{ack}
We thank Jens Hjorth, Steen H. Hansen, Priya Natarajan and the anonymous referee for helpful comments and
discussions. The Dark Cosmology Centre is funded by the Danish
National Research Foundation.  This work was supported by the European
Community's Sixth Framework Marie Curie Research Training Network
Programme, Contract No.MRTN-CT-2004-505183 "ANGLES".  This research was supported in part by the National Science Foundation under Grant No. PHY99-07949. 
\end{ack}

{}

\label{lastpage}

\end{document}